\documentclass[aip, rsi, reprint, 10pt,superscriptaddress]{revtex4-1}
\usepackage{graphicx}

\begin{document}
	
\title {An ultra-compact low temperature scanning probe microscope for magnetic fields above 30~T}
\author{L. Rossi}
\affiliation{High Field Magnet Laboratory (HFML–-EMFL), Radboud University, Nijmegen, Netherlands}
\affiliation{Radboud University, Institute of Molecules and Materials, Nijmegen, Netherlands}
\author{J. W. Gerritsen}
\affiliation{Radboud University, Institute of Molecules and Materials, Nijmegen, Netherlands}
\author{L. Nelemans}
\affiliation{High Field Magnet Laboratory (HFML–-EMFL), Radboud University, Nijmegen, Netherlands}
\author{A. A. Khajetoorians}
\affiliation{Radboud University, Institute of Molecules and Materials, Nijmegen, Netherlands}
\author{B. Bryant}
\email{ben.bryant@ru.nl}
\affiliation{High Field Magnet Laboratory (HFML–-EMFL), Radboud University, Nijmegen, Netherlands}
\affiliation{Radboud University, Institute of Molecules and Materials, Nijmegen, Netherlands}
\date{\today}
	
\begin{abstract}

We present the design of a highly compact High Field Scanning Probe Microscope (HF-SPM) for operation at cryogenic temperatures in an extremely high magnetic field, provided by a water-cooled Bitter magnet able to reach 38~T. The HF-SPM is 14 mm in diameter: an Attocube nano-positioner controls the coarse approach of a piezo resistive AFM cantilever to a scanned sample. The Bitter magnet constitutes an extreme environment for SPM due to the high level of vibrational noise;  the Bitter magnet noise at frequencies up to 300~kHz is characterized and noise mitigation methods are described.  The performance of the HF-SPM is demonstrated by topographic imaging and noise measurements at up to 30~T. Additionally, the use of the SPM as a three-dimensional dilatometer for magnetostriction measurements is demonstrated via measurements on a magnetically frustrated spinel sample. 
	
\end{abstract}
\maketitle

\section{Introduction}

Scanning Probe Microscopy (SPM) provides a versatile tool for high-resolution imaging. It allows spatial features of materials which differ from the bulk by their magnetic, electronic or other properties to be probed down to nanometer length scales. Probes have been developed for Atomic Force Microscopy (AFM) and related SPM techniques at cryogenic temperatures and in magnetic fields\cite{Kundhikanjana2011,Geng2014a,Zhou2014,Galvis2015}, which have enabled novel properties of magnetic and multiferroic domain walls to be explored\cite{Ma2015_2,Geng2014}. Magnetic Force Microscopy (MFM) has been used to image spatially inhomogeneous magnetic textures such as skyrmions\cite{Milde2013} and bubble domains \cite{Wulferding2017, Jeong2015}, and can also be used to image phase coexistence at metamagnetic phase transitions\cite{Zhou2015, Sheng2018}. However, since almost all SPM setups to date have utilized laboratory superconducting magnets, the maximum field has been limited to 20~T\cite{Zhou2014}. This leaves many field-induced phase transitions, such as in frustrated spinels\cite{Ueda2005} and multiferroics\cite{Tokunaga2011}, out of reach. Static fields above 30 T can be attained in high-field facilities by water-cooled, resistive Bitter magnets, or hybrid resistive-superconducting magnets. Previously, we demonstrated a Scanning Tunneling Microscope (STM) capable of operating in a water-cooled Bitter Magnet at up to 34~T \cite{Tao2017}. Here we present the extension of this concept to a versatile High Field Scanning Probe Microscope (HF-SPM) which can operate at up to 38~T. The SPM can be used in a variety of dynamic modes at frequencies up to 300~kHz, enabling a wider range of experiments compared to STM, however it also brings additional challenges for high-field operation since high-frequency noise from the Bitter magnet must be considered and its effect minimized. In this paper we demonstrate static and dynamic Atomic Force Microscopy imaging and Magnetic Force Microscopy at fields up to 30 T.

\section{Instrument design}

\begin{figure}[b]
\begin{centering}
\includegraphics[scale=1]{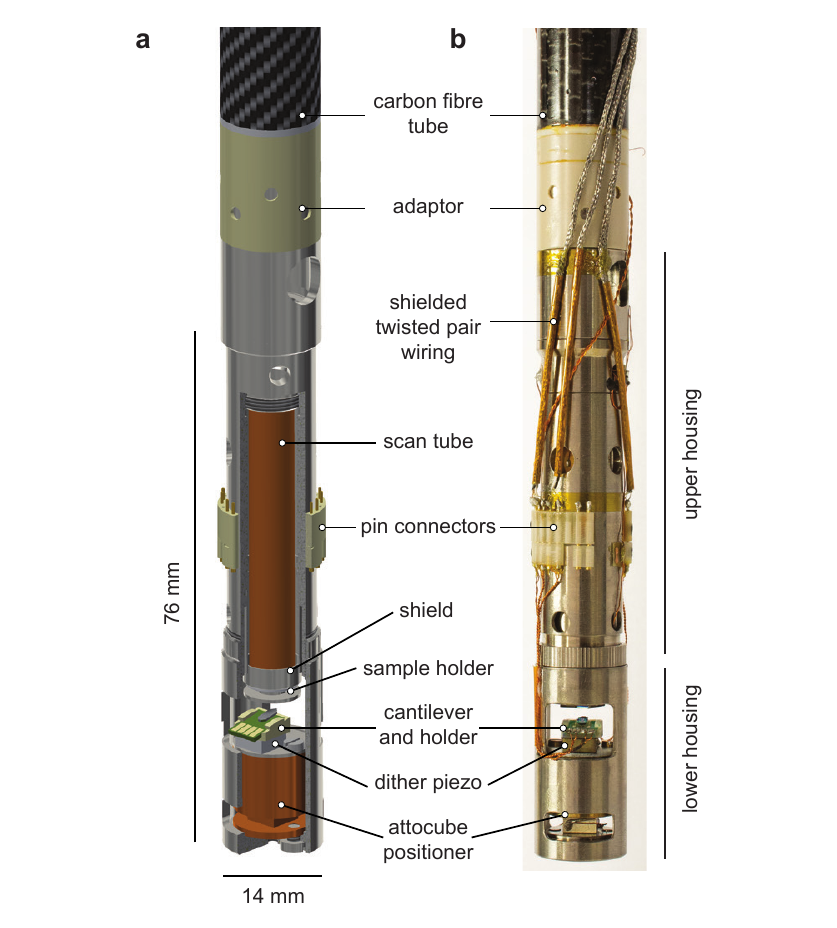}
\caption{(color online). (a) diagram and (b) image of the HF-SPM head, with the main components indicated. See text for details.}
\label{FigureA}
\end{centering}
\end{figure}

A suite of Bitter magnets are available at the High Field Magnet Laboratory (HFML) in Nijmegen, including two 38~T magnets each with a 32~mm room temperature bore\cite{Wiegers2012}, and a 30~T magnet with a 50~mm bore\cite{Wiegers2008}. The HF-SPM is specially designed to work in the extreme conditions of the Bitter magnet: as well as in high magnetic field and at cryogenic temperatures, the SPM has to operate within the high-noise environment produced by the turbulent magnet cooling water. For the 38~T magnets, inside the cryostat helium space the bore is 24~mm, and the inner diameter of the sample vacuum tube is 20~mm. The SPM has to be compact enough to fit within this space, and it should be as rigid as possible to make it insensitive to vibration.

Figure \ref{FigureA} shows the layout of the HF-SPM head. The design is inspired by the previous high field STM \cite{Tao2017}, with a compact, concentric cylindrical arrangement, which gives a high degree of rigidity. The whole SPM head measures 76~mm in length and its diameter is 14~mm, providing some clearance inside the 20~mm vacuum tube to prevent the transmission of vibrational noise. The housing of the HF-SPM is made of nonmagnetic grade 2 Titanium. The head is split into a lower housing, which contains the cantilever and coarse approach mechanism, and an upper housing, which contains the piezo scan tube, onto which the sample holder is mounted. The lower housing screws onto the upper and is secured with a locking nut, it is removable for easier sample and cantilever exchange. The HF-SPM is mounted via an insulating adaptor to a carbon fibre tube, which is light and relatively stiff, and provides some degree of vibration damping. 

\subsection{Cantilever}

Because of the restricted dimensions of the HF-SPM, it is not practical to use an optical beam deflection system to measure the cantilever deflection. A quartz tuning-fork or needle-type sensor is also not feasible since these ideally require a low-temperature amplifier \cite{Giessibl2011}, which is ruled out by the high magnetic field and space constraints. Therefore, we opted for self-sensing silicon cantilevers, in which a piezoresistor is integrated into the base of the cantilever; the resistance change is used to measure the deflection \cite{tortonese1993}. We use cantilevers produced by Hitachi, namely PRC120 and PRC400 with resonance frequencies of $\approx$ 300~kHz and 43~kHz, and spring constants of 30-40~N/m and 2-4~N/m, respectively\cite{Takahashi2002}. We have also employed piezoresitive cantilevers from SCL-Sensor Tech\cite{Dukic2015}. The cantilevers have additional reference piezoresistors, of the same material as the cantilever piezoresistor, integrated into the chip. The deflection is measured using a Wheatstone bridge setup: since the cantilever and the reference have approximately the same temperature coefficient of resistivity and magnetoresistance, the measurement is relatively insensitive to changes in temperature and magnetic field. For MFM measurements we coat the cantilevers by electron-beam evaporation with 2~nm of Ti as adhesion layer, 20 to 40~nm of Co, and 2~nm of Au to prevent the oxidation of the Co. The Hitachi cantilevers are mounted in a holder, 3D printed in alumina. The cantilever PCB is held by a Cu-Be spring and slides in or out of the holder, making for easy cantilever exchange. To drive the oscillation the holder is mounted on a dither piezo (PL055.30, PICeramic). 

\subsection{Coarse approach \& sample scanner }

The cantilever holder and dither piezo are mounted on an 11~mm diameter ANPz30 Attocube nano-positioner, which allows the coarse approach of the cantilever to the sample. It is reliable both at room and cryogenic temperatures. The Attocube positioner provides for a modular design that makes it easy not only to change the components in case of need, but also provides flexibility to employ different cantilever holders.

The upper housing contains the piezo scan tube; the SPM uses a scanned sample design. A 40mm long tube (EBL \# 3, EBL products) is employed, which allows at least a 5 x 5 $\mu$m scan area at 4 K at $\pm$ 140 V. The scan tube is isolated at each end by Macor spacers. A grounded shield at the bottom (sample) end of the scan tube shields the sample from the scan tube voltages. The sample holder is screwed into the end of the tube and is isolated from the shield. A separate electrical connection is provided to the sample holder, to enable conductive AFM or STM modes. To reduce vibrational noise sensitivity the assembly at the sample end of the scan tube is designed to be as light as possible, therefore the shield and sample holder are constructed of titanium; the sample holder weighs less than 0.2~g. 

On the upper housing are pin connectors (plugs and sockets) for the wires to the lower housing; the connections to the piezoresitive cantilever, the dither piezo and the Attocube positioner. The upper pin connectors are machined from Macor, the lower connectors are 3D printed in plastic. For sample exchange, the pin connectors are unplugged and the lower housing is removed: the sample can then be unscrewed from the scan tube. The upper housing remains connected to the carbon fibre mounting tube. 

\subsection{Wiring \& Electronics}

Above the pin connectors on the SPM upper housing, the cantilever and dither signals use shielded twisted pair cables (GVLZ141, GVL Cryoenginneering): by using a balanced signal, this allows to reduce electronic noise and crosstalk. The sample and tip connections use miniature coaxial cables (Type C, Lakeshore Cryotronics). The piezoresistive cantilever and its reference are connected via the shielded cables to a home-made purpose-built preamplifier based around an INA 217 Instrumentation amplifier (Texas Instruments). The amplifier gives 1000 x gain to the Wheatstone bridge bias at 100~kHz bandwidth, or 100 x gain at 1 MHz. The bias across the bridge is set at 0.8~V to reduce ohmic heating by the piezoresistors to less than 0.1~K, measured on the SPM body. The SPM is controlled by a NANONIS system (SPECS GmbH), which allows many SPM modes to be employed, including both amplitude and frequency modulation, and dual-pass modes for MFM.

\subsection{Vacuum system, Cryostat and Vibration isolation}

\begin{figure}
\begin{centering}
\includegraphics[scale=1]{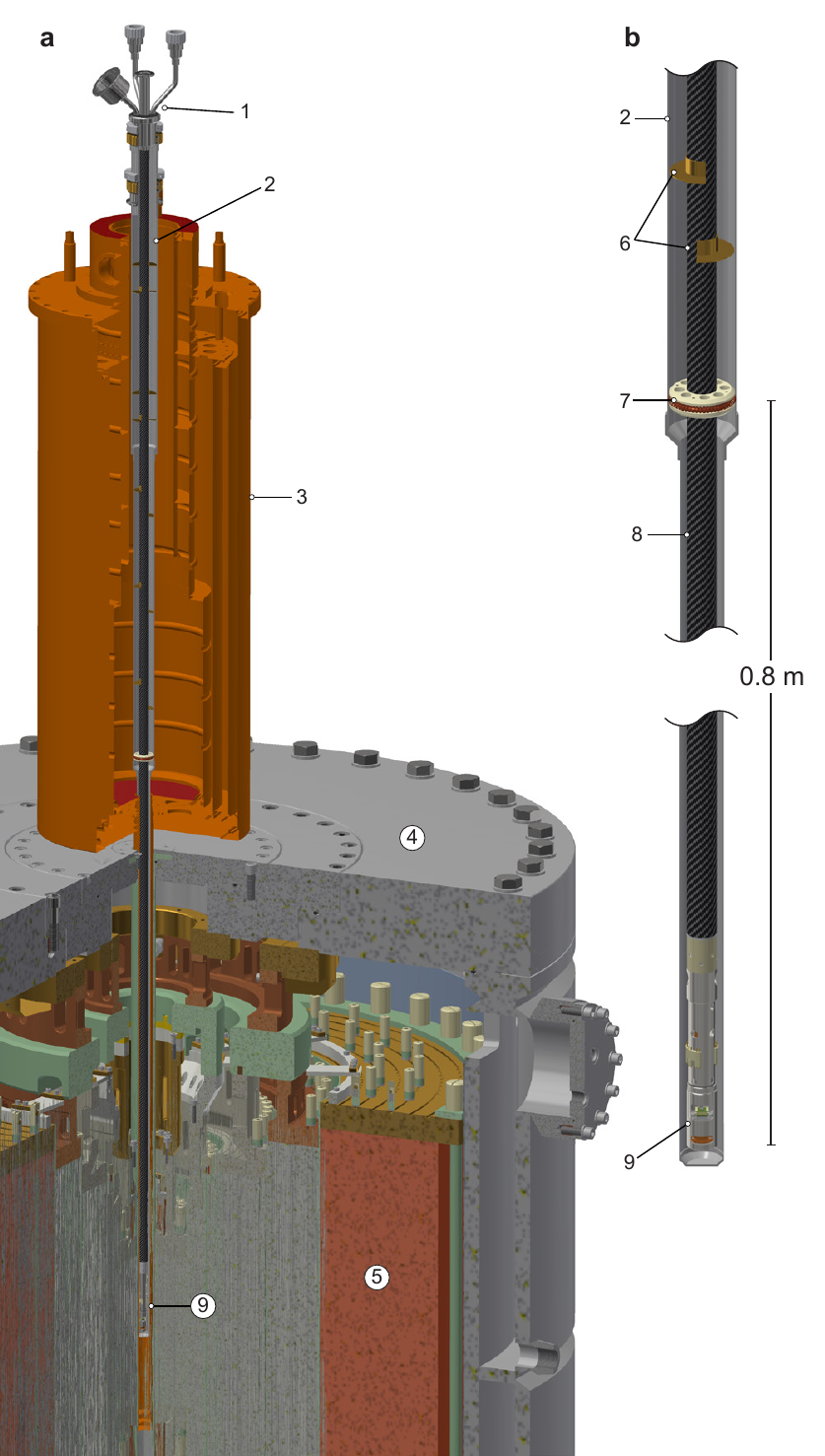}
\caption{(color online). (a) Drawing of the HF-SPM assembly, cryostat and Bitter magnet. (b) Detail of the HF-SPM assembly in vacuum tube. (1) Top assembly with vacuum feedthroughs (2) vacuum tube (3) helium bath cryostat:  support structure not shown (4) housing of 38 T magnet (5) magnet Bitter disks (6) thermal baffles (7) centering ring with Cu-Be springs (8) carbon fibre mounting tube (9) SPM head. The carbon tube is free from the vacuum tube over a 0.8~m length, affording some integral vibration isolation.}
\label{FigureB}
\end{centering}
\end{figure}

Figure \ref{FigureB}a shows the setup of the HF-SPM and helium bath cryostat in one of the 38~T Bitter magnets at HFML. Inside the cryostat, the vacuum tube carrying the SPM is vibrationally decoupled from the helium bath by a ring of Cu-Be springs.  Inside the vacuum tube a centering ring with Cu-Be springs is mounted around 0.8~m above the SPM head (figure \ref{FigureB}b), giving this length of carbon fibre tube as a damped pendulum to aid in vibration isolation. The external vibration isolation system consists of a support structure and active damping units: it is not shown in figure \ref{FigureB} since it is described in more detail in the context of the high field STM at HFML\cite{Tao2017}. The support structure for the cryostat is fixed to the floor separately from the Bitter magnet: for the 38~T magnets this structure has been optimized to make it as rigid as possible. The cryostat is mounted on the support structure via two active damping bars, the support can be adjusted so that the tail of the cryostat does not touch the magnet.

Vacuum feedthroughs for the piezo voltages and cantilever connections are fitted at the top of the vacuum tube: the pre-amplifier is directly mounted onto the vacuum feedthrough. For cryogenic operation the vacuum tube is evacuated to a pressure of better than $10^{-5}$~mbar, before inserting it into the cryostat. The SPM is normally operated at 4.3~K, in He exchange gas at a pressure of around $10^{-2}$~mbar: by pumping on the cryostat helium bath, it is possible to reach 1.35~K at a similar exchange gas pressure. 

\section{Characterization and mitigation of Bitter Magnet Noise}

\begin{figure}
\begin{centering}
\includegraphics[scale=1]{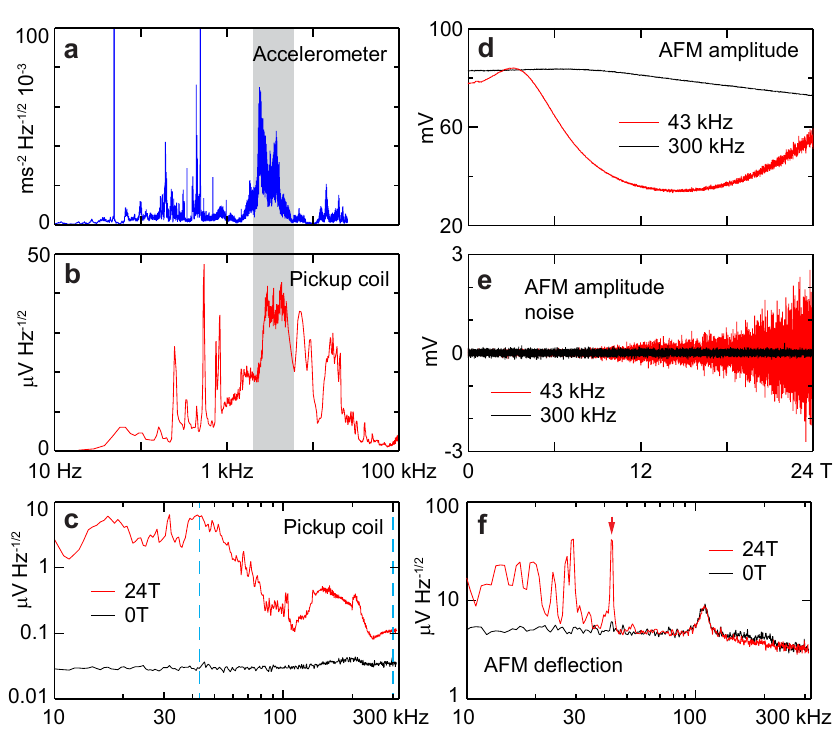}
\caption{(color online). Noise characterization of the 30~T Bitter magnet. (a) Vibration FFT spectrum recorded on the top plate of the magnet at 24~T, with 140~L/s water flow. The frequency band from 2 to 6~kHz is highlighted. (b) Magnetic field noise FFT spectrum recorded simultaneously with (a) using a pickup coil in the field center. (c) Pickup coil FFT spectra at zero field and at 24~T. An aluminum tube is employed to shield some of the AC component of the magnetic field. The frequencies of the AFM cantilevers employed in the HF-SPM  (43 and 300~kHz) are indicated. (d) AFM cantilever deflection signal in the 30 T magnet, as a function of field, for cantilevers with 43 and 300~kHz resonant frequencies. (e) Noise on AFM cantilever deflection signal in (d). (f) FFT spectrum of the undriven 43 kHz cantilever deflection at zero and 24~T. The resonant frequency is indicated.}
\label{FigureC}
\end{centering}
\end{figure}

In order to effect a reasonable signal-to-noise level for SPM in a Bitter magnet, it was necessary to characterize the noise produced by the magnet at a wide range of frequencies. To this end, we measured both the vibrational spectrum on the Bitter magnet housing with accelerometers, and the AC component of the magnetic field with a pickup coil. Two Endevco Isotron accelerometers of 1~kHz and 30~kHz bandwidth were placed on top of the magnet housing. A calibrated pickup coil with 289 turns was placed in the field center: the coil voltage was amplified by a 100 x gain amplifier with 400~kHz bandwidth. Figure \ref{FigureC} shows the results for the 30~T, 50~mm bore magnet at HFML\cite{Wiegers2008}. The vibration FFT spectrum is shown in \ref{FigureC}a: the magnet is energized to 24~T with a cooling water flow of 140 L/s. A wide band of vibrational peaks is seen in the range 2 to 6~kHz, which contain a large proportion of the integrated intensity: almost all the vibrational energy is produced at frequencies above 2~kHz. This noise derives from the turbulent flow of the cooling water through the magnet: these measurements are in line with previous measurements on Bitter magnets with the largest vibrational peak at 3.4~kHz \cite{Soghomonian2000}. Figure \ref{FigureC}b shows an FFT spectrum of the magnetic field noise recorded using the pickup coil, simultaneously with the vibration measurements. Several sharp peaks are seen below 1~kHz: these are the odd harmonics of 50~Hz and can be attributed to the field ripple from the DC power supply\cite{Soghomonian2000}. The field ripple up to 1 kHz amounts to 0.5 ppm at 24 T. At higher frequencies we see a similar  band of vibrational peaks in the 2 to 6~kHz range as seen in the accelerometer data, and a further band 10 to 22~kHz. These noise peaks can also be attributed to the cooling water flow: this causes the magnet and coil to vibrate relative to each other. Since the field is relatively inhomogeneous this produces an effective AC component of the magnetic field. The higher frequencies are more pronounced in the pickup coil spectrum than in the vibration data: this may indicates that the magnet does not vibrate as a single solid mass, but that the motion of the more massive outer coils, which will vibrate at lower frequencies, dominate the vibrational spectrum measured on the magnet housing.

We now discuss some methods for mitigating the effect of the Bitter magnet noise: we will focus on the effects on SPM measurements, but the analysis is relevant to any measurement technique which is sensitive to the AC component of the field or to vibration.  In a low-temperature experiment in a Bitter magnet a metallic cryostat will act as an eddy current shield to the AC component of the field: the effect of this inductive shielding has been discussed in detail\cite{Meyer1989}. To provide a similar degree of screening in room-temperature measurements an aluminum shielding tube (50~mm OD, 34~mm ID) was inserted in the magnet bore. Figure \ref{FigureC}c shows the magnetic field noise recorded using the pickup coil at zero field and 24~T, with 140~L/s water flow, with the aluminum shielding tube in place. The shielding tube is highly effective at reducing the noise pickup, particularly at frequencies $>$ 50~kHz, however considerable field-induced noise remains. At 43~kHz there is a 200-fold increase in noise pickup from zero to 24 T: this sharply reduces at frequencies above 200~kHz to a less than threefold increase at 300~kHz. 

Figure \ref{FigureC}d-f shows the effect of the magnetic field noise on the HF-SPM cantilever deflection signal. The SPM is mounted in the 30~T magnet, at room temperature and in vacuum ($\approx10^{-2}$~mbar), with the aluminum shielding tube in place and 140~L/s of cooling water. The AFM cantilever deflection signal is shown in figure \ref{FigureC}e as a function of field, for cantilevers with 43 and 300~kHz resonant frequencies: PLL feedback is used to drive the cantilever at its resonant frequency. The noise on this signal is shown in figure \ref{FigureC}e. In the case of the 43~kHz cantilever a large increase in noise is seen, from around 40~$\mu$V rms at low field to around 1~mV rms at 24~T. For the 300~kHz cantilever however no increase in noise is seen, mirroring the effect seen in the pickup coil in figure \ref{FigureC}c. FFT spectra of the 43~kHz cantilever deflection (\ref{FigureC}f) show that the additional noise with field in the cantilever deflection is sharply peaked at the resonant frequency, suggesting that the cantilever is primarily sensitive to vibration rather than directly to electrical pickup. This implies that the SPM is being induced to vibrate by the effective AC component of the magnetic field, effectively making a short-cut around the vibration isolation: since the SPM is nonmagnetic the likely coupling mechanism is via eddy currents induced in the titanium body of the SPM itself. At room temperature the skin depth for Ti at 43~kHz is 1.5~mm, which is comparable to the wall thickness of the SPM body, suggesting that this coupling mechanism can be effective. The best way to avoid this noise coupling would be to construct the SPM head entirely from an insulating material: for the present design though, operating at a cantilever resonant frequency higher than 200~kHz allows us to reduce the field-induced noise to an acceptable level, enabling SPM operation in the Bitter magnet.

\section{Imaging tests with the HF-SPM}

\begin{figure}
\begin{centering}
\includegraphics[scale=1]{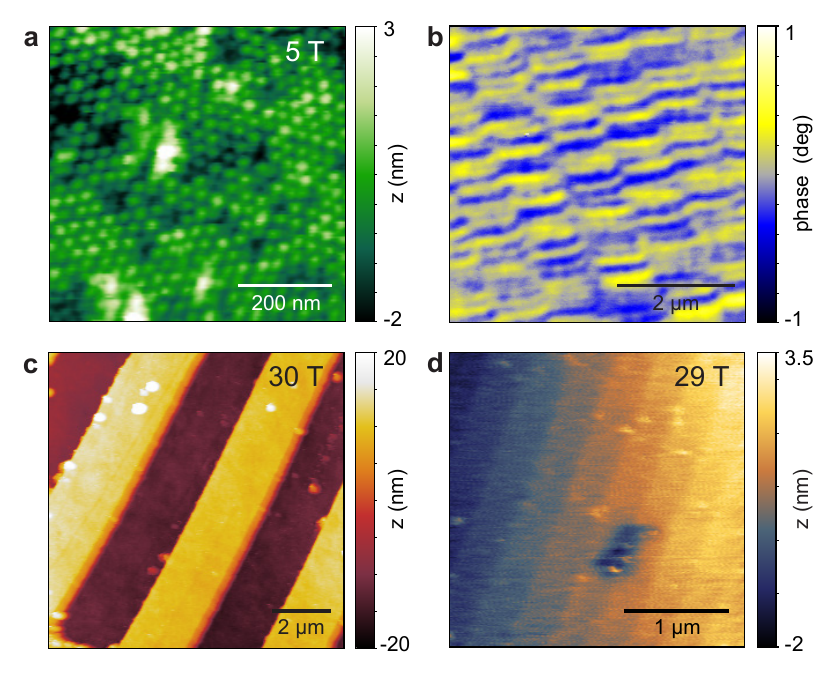}
\caption{(color online). Imaging tests with the HF-SPM. (a) 800 x 800~nm AFM topograph of an array of CoFe$_2$O$_4$ spinel nanodots, imaged at 4.3~K and 5~T in a superconducting magnet. (b) 5 x 5~$\mu$m MFM image of a hard disk test sample, imaged at 4.3~K at zero field, with 30~nm lift height, using a Co-coated tip. (c) 10 x 10~$\mu$m AFM topograph of a SiO$_2$ calibration grid, imaged at room temperature at 30~T in the 38~T Bitter magnet. (d) 2.8 x 2.8~$\mu$m AFM topograph of a single crystal CdCr$_2$O$_4$ spinel sample, imaged at 4.3~K at 29~T in the 30~T magnet. Atomic terraces are visible, with 0.5~nm high steps.}
\label{FigureD}
\end{centering}
\end{figure}

In order to verify the capabilities of the HF-SPM we performed some imaging tests under different conditions. The SPM is operated in tapping (constant amplitude) mode. Initially we tested the instrument inside a superconducting magnet. Figure \ref{FigureD}(a) shows an image of an array of 30~nm diameter spinel CoFe$_2$O$_4$ nanodots, obtained with a PRC400 cantilever at a magnetic field of 5~T, at 4.3~K. This image was obtained without any external vibration isolation: the superconducting magnet is situated on the first floor of the laboratory, with a high level of background vibrational noise. The high quality of the image demonstrates that the AFM is relatively insensitive to external vibration. We also performed MFM imaging at low temperature using the PRC120 tips to ensure that these relatively stiff cantilevers provide sufficient force sensitivity. Figure \ref{FigureD}(b) shows a phase MFM image recorded on an hard disk test sample at 4.3~K and zero field, at a lift height of 30~nm.  We used a Co coated PRC120 cantilever with 300~kHz resonance frequency and a \emph{Q} factor of 5000. The phase signal to noise is around 60: it is also possible to perform MFM imaging using a phase locked loop (PLL) to measure a frequency-shift image, with comparable signal-to-noise. 

Finally, we tested the performance of the SPM in high magnetic field. Figure \ref{FigureD}(c) is a tapping mode image of a SiO$_2$ calibration grid collected at 30~T, at room temperature, in the 38~T Bitter magnet, with 140~L/s of water flow. The image was recorded at $10^{-2}$~mbar using a 300~kHz cantilever. Detailed images of the SiO$_2$ test grid allow us to demonstrate a lateral resolution better than 20~nm and a \emph{z} resolution of 0.13~nm. The \emph{z} resolution at high field is thus comparable to a commercial AFM. As with the high-field STM \cite{Tao2017}, the \emph{x-y} resolution is likely limited by the vibrational modes of the 40~mm long scan tube, but it is sufficient for scan modes such as MFM. Figure \ref{FigureD}(d) shows a tapping mode image of the [111] surface of a single crystal sample of spinel CdCr$_2$O$_4$, taken at 4.3~K at 29~T in the 30~T Bitter magnet, with 140~L/s of water flow. The vibration isolation for this magnet is not yet fully optimized, hence  the \emph{z} resolution here is 0.3~nm. Nonetheless, atomic terraces with 0.5~nm high steps are visible.

\section{Dilatometry with the HF-SPM}

\begin{figure}[b]
\begin{centering}
\includegraphics[scale=1]{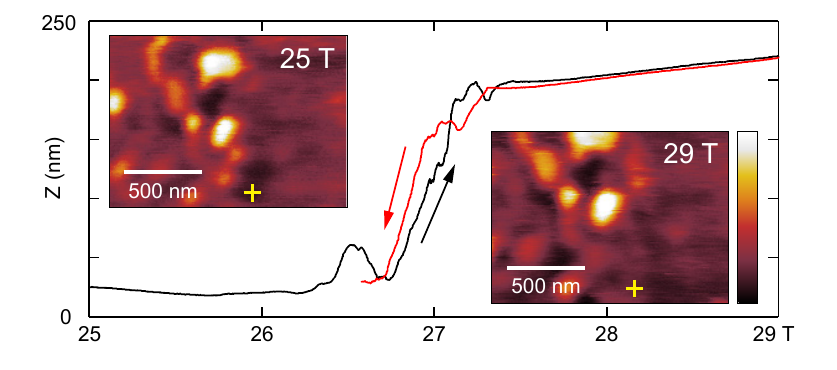}
\caption{(color online). Magnetostriction \emph{z} measurement of CdCr$_2$O$_4$ with the HF-SPM, from 25 to 29~T at 5.4~K. Contact mode topographic images at the same scan coordinates at 25 and 29~T are shown: the same color scale of 9~nm is used in both images, yellow crosses indicate the position of the tip during the magnetostriction measurement.}
\label{FigureE}
\end{centering}
\end{figure}

The HF-SPM can also be used as a dilatometer to measure magnetostriction. In this case the AFM is operated in constant-force contact mode by monitoring the static deflection of the cantilever.  A magnetostriction measurement is performed by first putting the tip in contact with the sample and keeping the feedback loop on while the field is ramped. The magnetostriction of the sample can be measured directly as the change in \emph{z}. Figure \ref{FigureE} shows such a magnetostriction measurement of the magnetically frustrated spinel CdCr$_2$O$_4$. This compound is known to have a magnetostructural transition at low temperature at 27~T, from a low-field antiferromagnetic, tetragonal (c $>$ a $=$ b) state to a high-field ferrimagnetic, cubic phase \cite{Ueda2005,Inami2006}. The HF-AFM was operated in the 38~T Bitter magnet with 130~L/s of water flow, at 5.4~K. Single crystal samples oriented in the [111] direction are used: the field is ramped from 25 to 29~T. The sample is observed to expand along the [111] direction by 175~nm at 27~T. The transition is broad and hysteretic: it shows peaks and dips, possibly related to the domain structure of the sample in the tetragonal phase.

Using piezoresistive cantilevers to measure magnetostriction has the advantage that the pressure applied to the sample is negligible compared to, e.g. a capacitative bulk dilatometer\cite{kuchler2017}. Piezoresistive cantilevers have been used to perform magnetostriction measurements previously but without any piezo positioning of the sample\cite{Park2009}. Using the HF-SPM for this type of measurements allows to us to take advantage of the calibrated scan piezo to collect more accurate data. 

Performing dilatometry with an SPM also makes it possible to make three-dimensional magnetostriction measurements. This is demonstrated in Figure \ref{FigureE}, where contact-mode images of the CdCr$_2$O$_4$ sample, at the same scan coordinates, are shown at 25 and 29~T. There is a lateral offset, which at $\approx$ 170~nm is similar in magnitude to the change in \emph{z}. This is expected, since we are imaging the (111) face of the CdCr$_2$O$_4$ crystal, and at the transition to the high-field cubic phase the c-axis contracts and the a and b-axes expand\cite{Inami2006}. Using this technique magnetostriction in three axes can be measured with a single field sweep. Spatially-resolved measurements of magnetostriction could potentially be made using the SPM, providing a method for local measurements of magnetostriction in nanocomposites, e.g. FePt nanoparticles\cite{Reid2018} and nanopillars for multiferroic composites \cite{Poddar2018}.

\section{Conclusions}

In this paper we have determined that an effective strategy for mitigating field noise in measurements in Bitter magnets is to make lock-in measurements at high frequencies, above 200~kHz. Combined with effective vibrational isolation, this can make highly-sensitive measurements possible at up to 38~T. Clearly however, the ideal configuration for SPM, or any other vibrationally sensitive type of measurements, in a Bitter magnet would be for the instrument itself to be made of nonmagnetic and non-conductive material such as Macor or sapphire\cite{White2011,Trainer2017}, so as to avoid eddy current pickup in the measurement probe entirely. 

The HF-SPM has been designed to be highly compact and insensitive to vibration: these attributes make it a potentially useful design for applications other than in a Bitter magnet. The vibrational insensitivity of the SPM head means that complex external vibration damping is not required. The compact size allows for either higher magnetic field superconducting magnets to be employed, or for overall system size to be reduced.  The small mass of the SPM minimises the cryogenic cooling power required. The minimal bore diameter required for the HF-SPM could also enable operation in permanent magnet arrays reaching 2~T \cite{Tayler2017}. 

\section*{Acknowledgments}

The CoFe$_2$O$_4$ nanodot samples were supplied by Jin Xu (University of Groningen) and the CdCr$_2$O$_4$ single crystal samples by Hiroaki Ueda (Kyoto University). The authors would like to thank Frans Wijnen and Andries van Ouden (HFML) for useful discussions regarding the Bitter magnet noise. 

\bibliography{Mendeley}

\begin{thebibliography}{30}%
\makeatletter
\providecommand \@ifxundefined [1]{%
 \@ifx{#1\undefined}
}%
\providecommand \@ifnum [1]{%
 \ifnum #1\expandafter \@firstoftwo
 \else \expandafter \@secondoftwo
 \fi
}%
\providecommand \@ifx [1]{%
 \ifx #1\expandafter \@firstoftwo
 \else \expandafter \@secondoftwo
 \fi
}%
\providecommand \natexlab [1]{#1}%
\providecommand \enquote  [1]{``#1''}%
\providecommand \bibnamefont  [1]{#1}%
\providecommand \bibfnamefont [1]{#1}%
\providecommand \citenamefont [1]{#1}%
\providecommand \href@noop [0]{\@secondoftwo}%
\providecommand \href [0]{\begingroup \@sanitize@url \@href}%
\providecommand \@href[1]{\@@startlink{#1}\@@href}%
\providecommand \@@href[1]{\endgroup#1\@@endlink}%
\providecommand \@sanitize@url [0]{\catcode `\\12\catcode `\$12\catcode
  `\&12\catcode `\#12\catcode `\^12\catcode `\_12\catcode `\%12\relax}%
\providecommand \@@startlink[1]{}%
\providecommand \@@endlink[0]{}%
\providecommand \url  [0]{\begingroup\@sanitize@url \@url }%
\providecommand \@url [1]{\endgroup\@href {#1}{\urlprefix }}%
\providecommand \urlprefix  [0]{URL }%
\providecommand \Eprint [0]{\href }%
\providecommand \doibase [0]{http://dx.doi.org/}%
\providecommand \selectlanguage [0]{\@gobble}%
\providecommand \bibinfo  [0]{\@secondoftwo}%
\providecommand \bibfield  [0]{\@secondoftwo}%
\providecommand \translation [1]{[#1]}%
\providecommand \BibitemOpen [0]{}%
\providecommand \bibitemStop [0]{}%
\providecommand \bibitemNoStop [0]{.\EOS\space}%
\providecommand \EOS [0]{\spacefactor3000\relax}%
\providecommand \BibitemShut  [1]{\csname bibitem#1\endcsname}%
\let\auto@bib@innerbib\@empty
\bibitem [{\citenamefont {Kundhikanjana}\ \emph {et~al.}(2011)\citenamefont
  {Kundhikanjana}, \citenamefont {Lai}, \citenamefont {Kelly},\ and\
  \citenamefont {Shen}}]{Kundhikanjana2011}%
  \BibitemOpen
  \bibfield  {author} {\bibinfo {author} {\bibfnamefont {W.}~\bibnamefont
  {Kundhikanjana}}, \bibinfo {author} {\bibfnamefont {K.}~\bibnamefont {Lai}},
  \bibinfo {author} {\bibfnamefont {M.~A.}\ \bibnamefont {Kelly}}, \ and\
  \bibinfo {author} {\bibfnamefont {Z.-X.}\ \bibnamefont {Shen}},\ }\href
  {\doibase 10.1063/1.3554438} {\bibfield  {journal} {\bibinfo  {journal}
  {Review of Scientific Instruments}\ }\textbf {\bibinfo {volume} {82}},\
  \bibinfo {pages} {033705} (\bibinfo {year} {2011})}\BibitemShut {NoStop}%
\bibitem [{\citenamefont {Geng}\ and\ \citenamefont {Wu}(2014)}]{Geng2014a}%
  \BibitemOpen
  \bibfield  {author} {\bibinfo {author} {\bibfnamefont {Y.}~\bibnamefont
  {Geng}}\ and\ \bibinfo {author} {\bibfnamefont {W.}~\bibnamefont {Wu}},\
  }\href {\doibase 10.1063/1.4874006} {\bibfield  {journal} {\bibinfo
  {journal} {Review of Scientific Instruments}\ }\textbf {\bibinfo {volume}
  {85}},\ \bibinfo {pages} {053901} (\bibinfo {year} {2014})}\BibitemShut
  {NoStop}%
\bibitem [{\citenamefont {Zhou}\ \emph {et~al.}(2014)\citenamefont {Zhou},
  \citenamefont {Wang}, \citenamefont {Hou},\ and\ \citenamefont
  {Lu}}]{Zhou2014}%
  \BibitemOpen
  \bibfield  {author} {\bibinfo {author} {\bibfnamefont {H.}~\bibnamefont
  {Zhou}}, \bibinfo {author} {\bibfnamefont {Z.}~\bibnamefont {Wang}}, \bibinfo
  {author} {\bibfnamefont {Y.}~\bibnamefont {Hou}}, \ and\ \bibinfo {author}
  {\bibfnamefont {Q.}~\bibnamefont {Lu}},\ }\href {\doibase
  10.1016/j.ultramic.2014.07.011} {\bibfield  {journal} {\bibinfo  {journal}
  {Ultramicroscopy}\ }\textbf {\bibinfo {volume} {147}},\ \bibinfo {pages}
  {133} (\bibinfo {year} {2014})}\BibitemShut {NoStop}%
\bibitem [{\citenamefont {Galvis}\ \emph {et~al.}(2015)\citenamefont {Galvis},
  \citenamefont {Herrera}, \citenamefont {Guillam{\'{o}}n}, \citenamefont
  {Azpeitia}, \citenamefont {Luccas}, \citenamefont {Munuera}, \citenamefont
  {Cuenca}, \citenamefont {Higuera}, \citenamefont {D{\'{i}}az}, \citenamefont
  {Pazos}, \citenamefont {Garc{\'{i}}a-Hernandez}, \citenamefont
  {Buend{\'{i}}a}, \citenamefont {Vieira},\ and\ \citenamefont
  {Suderow}}]{Galvis2015}%
  \BibitemOpen
  \bibfield  {author} {\bibinfo {author} {\bibfnamefont {J.~A.}\ \bibnamefont
  {Galvis}}, \bibinfo {author} {\bibfnamefont {E.}~\bibnamefont {Herrera}},
  \bibinfo {author} {\bibfnamefont {I.}~\bibnamefont {Guillam{\'{o}}n}},
  \bibinfo {author} {\bibfnamefont {J.}~\bibnamefont {Azpeitia}}, \bibinfo
  {author} {\bibfnamefont {R.~F.}\ \bibnamefont {Luccas}}, \bibinfo {author}
  {\bibfnamefont {C.}~\bibnamefont {Munuera}}, \bibinfo {author} {\bibfnamefont
  {M.}~\bibnamefont {Cuenca}}, \bibinfo {author} {\bibfnamefont {J.~A.}\
  \bibnamefont {Higuera}}, \bibinfo {author} {\bibfnamefont {N.}~\bibnamefont
  {D{\'{i}}az}}, \bibinfo {author} {\bibfnamefont {M.}~\bibnamefont {Pazos}},
  \bibinfo {author} {\bibfnamefont {M.}~\bibnamefont {Garc{\'{i}}a-Hernandez}},
  \bibinfo {author} {\bibfnamefont {A.}~\bibnamefont {Buend{\'{i}}a}}, \bibinfo
  {author} {\bibfnamefont {S.}~\bibnamefont {Vieira}}, \ and\ \bibinfo {author}
  {\bibfnamefont {H.}~\bibnamefont {Suderow}},\ }\href {\doibase
  10.1063/1.4905531} {\bibfield  {journal} {\bibinfo  {journal} {Review of
  Scientific Instruments}\ }\textbf {\bibinfo {volume} {86}},\ \bibinfo {pages}
  {013706} (\bibinfo {year} {2015})}\BibitemShut {NoStop}%
\bibitem [{\citenamefont {Ma}\ \emph {et~al.}(2015)\citenamefont {Ma},
  \citenamefont {Cui}, \citenamefont {Ueda}, \citenamefont {Tang},
  \citenamefont {Chen}, \citenamefont {Tamura}, \citenamefont {Wu},
  \citenamefont {Fujioka}, \citenamefont {Tokura},\ and\ \citenamefont
  {Shen}}]{Ma2015_2}%
  \BibitemOpen
  \bibfield  {author} {\bibinfo {author} {\bibfnamefont {E.~Y.}\ \bibnamefont
  {Ma}}, \bibinfo {author} {\bibfnamefont {Y.-T.}\ \bibnamefont {Cui}},
  \bibinfo {author} {\bibfnamefont {K.}~\bibnamefont {Ueda}}, \bibinfo {author}
  {\bibfnamefont {S.}~\bibnamefont {Tang}}, \bibinfo {author} {\bibfnamefont
  {K.}~\bibnamefont {Chen}}, \bibinfo {author} {\bibfnamefont {N.}~\bibnamefont
  {Tamura}}, \bibinfo {author} {\bibfnamefont {P.~M.}\ \bibnamefont {Wu}},
  \bibinfo {author} {\bibfnamefont {J.}~\bibnamefont {Fujioka}}, \bibinfo
  {author} {\bibfnamefont {Y.}~\bibnamefont {Tokura}}, \ and\ \bibinfo {author}
  {\bibfnamefont {Z.-X.}\ \bibnamefont {Shen}},\ }\href {\doibase
  10.1126/science.aac8289} {\bibfield  {journal} {\bibinfo  {journal}
  {Science}\ }\textbf {\bibinfo {volume} {350}},\ \bibinfo {pages} {538}
  (\bibinfo {year} {2015})}\BibitemShut {NoStop}%
\bibitem [{\citenamefont {Geng}\ \emph {et~al.}(2014)\citenamefont {Geng},
  \citenamefont {Das}, \citenamefont {Wysocki}, \citenamefont {Wang},
  \citenamefont {Cheong}, \citenamefont {Mostovoy}, \citenamefont {Fennie},\
  and\ \citenamefont {Wu}}]{Geng2014}%
  \BibitemOpen
  \bibfield  {author} {\bibinfo {author} {\bibfnamefont {Y.}~\bibnamefont
  {Geng}}, \bibinfo {author} {\bibfnamefont {H.}~\bibnamefont {Das}}, \bibinfo
  {author} {\bibfnamefont {A.~L.}\ \bibnamefont {Wysocki}}, \bibinfo {author}
  {\bibfnamefont {X.}~\bibnamefont {Wang}}, \bibinfo {author} {\bibfnamefont
  {S.-W.}\ \bibnamefont {Cheong}}, \bibinfo {author} {\bibfnamefont
  {M.}~\bibnamefont {Mostovoy}}, \bibinfo {author} {\bibfnamefont {C.~J.}\
  \bibnamefont {Fennie}}, \ and\ \bibinfo {author} {\bibfnamefont
  {W.}~\bibnamefont {Wu}},\ }\href {\doibase 10.1038/nmat3813} {\bibfield
  {journal} {\bibinfo  {journal} {Nature Materials}\ }\textbf {\bibinfo
  {volume} {13}},\ \bibinfo {pages} {163} (\bibinfo {year} {2014})}\BibitemShut
  {NoStop}%
\bibitem [{\citenamefont {Milde}\ \emph {et~al.}(2013)\citenamefont {Milde},
  \citenamefont {Kohler}, \citenamefont {Seidel}, \citenamefont {Eng},
  \citenamefont {Bauer}, \citenamefont {Chacon}, \citenamefont {Kindervater},
  \citenamefont {Muhlbauer}, \citenamefont {Pfleiderer}, \citenamefont
  {Buhrandt}, \citenamefont {Schutte},\ and\ \citenamefont
  {Rosch}}]{Milde2013}%
  \BibitemOpen
  \bibfield  {author} {\bibinfo {author} {\bibfnamefont {P.}~\bibnamefont
  {Milde}}, \bibinfo {author} {\bibfnamefont {D.}~\bibnamefont {Kohler}},
  \bibinfo {author} {\bibfnamefont {J.}~\bibnamefont {Seidel}}, \bibinfo
  {author} {\bibfnamefont {L.~M.}\ \bibnamefont {Eng}}, \bibinfo {author}
  {\bibfnamefont {A.}~\bibnamefont {Bauer}}, \bibinfo {author} {\bibfnamefont
  {A.}~\bibnamefont {Chacon}}, \bibinfo {author} {\bibfnamefont
  {J.}~\bibnamefont {Kindervater}}, \bibinfo {author} {\bibfnamefont
  {S.}~\bibnamefont {Muhlbauer}}, \bibinfo {author} {\bibfnamefont
  {C.}~\bibnamefont {Pfleiderer}}, \bibinfo {author} {\bibfnamefont
  {S.}~\bibnamefont {Buhrandt}}, \bibinfo {author} {\bibfnamefont
  {C.}~\bibnamefont {Schutte}}, \ and\ \bibinfo {author} {\bibfnamefont
  {A.}~\bibnamefont {Rosch}},\ }\href {\doibase 10.1126/science.1234657}
  {\bibfield  {journal} {\bibinfo  {journal} {Science}\ }\textbf {\bibinfo
  {volume} {340}},\ \bibinfo {pages} {1076} (\bibinfo {year}
  {2013})}\BibitemShut {NoStop}%
\bibitem [{\citenamefont {Wulferding}\ \emph {et~al.}(2017)\citenamefont
  {Wulferding}, \citenamefont {Kim}, \citenamefont {Yang}, \citenamefont
  {Jeong}, \citenamefont {Barros}, \citenamefont {Kato}, \citenamefont
  {Martin}, \citenamefont {Ayala-Valenzuela}, \citenamefont {Lee},
  \citenamefont {Choi}, \citenamefont {Ronning}, \citenamefont {Civale},
  \citenamefont {Baumbach}, \citenamefont {Bauer}, \citenamefont {Thompson},
  \citenamefont {Movshovich},\ and\ \citenamefont {Kim}}]{Wulferding2017}%
  \BibitemOpen
  \bibfield  {author} {\bibinfo {author} {\bibfnamefont {D.}~\bibnamefont
  {Wulferding}}, \bibinfo {author} {\bibfnamefont {H.}~\bibnamefont {Kim}},
  \bibinfo {author} {\bibfnamefont {I.}~\bibnamefont {Yang}}, \bibinfo {author}
  {\bibfnamefont {J.}~\bibnamefont {Jeong}}, \bibinfo {author} {\bibfnamefont
  {K.}~\bibnamefont {Barros}}, \bibinfo {author} {\bibfnamefont
  {Y.}~\bibnamefont {Kato}}, \bibinfo {author} {\bibfnamefont {I.}~\bibnamefont
  {Martin}}, \bibinfo {author} {\bibfnamefont {O.~E.}\ \bibnamefont
  {Ayala-Valenzuela}}, \bibinfo {author} {\bibfnamefont {M.}~\bibnamefont
  {Lee}}, \bibinfo {author} {\bibfnamefont {H.~C.}\ \bibnamefont {Choi}},
  \bibinfo {author} {\bibfnamefont {F.}~\bibnamefont {Ronning}}, \bibinfo
  {author} {\bibfnamefont {L.}~\bibnamefont {Civale}}, \bibinfo {author}
  {\bibfnamefont {R.~E.}\ \bibnamefont {Baumbach}}, \bibinfo {author}
  {\bibfnamefont {E.~D.}\ \bibnamefont {Bauer}}, \bibinfo {author}
  {\bibfnamefont {J.~D.}\ \bibnamefont {Thompson}}, \bibinfo {author}
  {\bibfnamefont {R.}~\bibnamefont {Movshovich}}, \ and\ \bibinfo {author}
  {\bibfnamefont {J.}~\bibnamefont {Kim}},\ }\href {\doibase 10.1038/srep46296}
  {\bibfield  {journal} {\bibinfo  {journal} {Scientific Reports}\ }\textbf
  {\bibinfo {volume} {7}},\ \bibinfo {pages} {1} (\bibinfo {year}
  {2017})}\BibitemShut {NoStop}%
\bibitem [{\citenamefont {Jeong}\ \emph {et~al.}(2015)\citenamefont {Jeong},
  \citenamefont {Yang}, \citenamefont {Yang}, \citenamefont {Ayala-Valenzuela},
  \citenamefont {Wulferding}, \citenamefont {Zhou}, \citenamefont {Goodenough},
  \citenamefont {de~Lozanne}, \citenamefont {Mitchell}, \citenamefont {Leon},
  \citenamefont {Movshovich}, \citenamefont {Jeong}, \citenamefont {Yeom},\
  and\ \citenamefont {Kim}}]{Jeong2015}%
  \BibitemOpen
  \bibfield  {author} {\bibinfo {author} {\bibfnamefont {J.}~\bibnamefont
  {Jeong}}, \bibinfo {author} {\bibfnamefont {I.}~\bibnamefont {Yang}},
  \bibinfo {author} {\bibfnamefont {J.}~\bibnamefont {Yang}}, \bibinfo {author}
  {\bibfnamefont {O.~E.}\ \bibnamefont {Ayala-Valenzuela}}, \bibinfo {author}
  {\bibfnamefont {D.}~\bibnamefont {Wulferding}}, \bibinfo {author}
  {\bibfnamefont {J.-S.}\ \bibnamefont {Zhou}}, \bibinfo {author}
  {\bibfnamefont {J.~B.}\ \bibnamefont {Goodenough}}, \bibinfo {author}
  {\bibfnamefont {A.}~\bibnamefont {de~Lozanne}}, \bibinfo {author}
  {\bibfnamefont {J.~F.}\ \bibnamefont {Mitchell}}, \bibinfo {author}
  {\bibfnamefont {N.}~\bibnamefont {Leon}}, \bibinfo {author} {\bibfnamefont
  {R.}~\bibnamefont {Movshovich}}, \bibinfo {author} {\bibfnamefont {Y.~H.}\
  \bibnamefont {Jeong}}, \bibinfo {author} {\bibfnamefont {H.~W.}\ \bibnamefont
  {Yeom}}, \ and\ \bibinfo {author} {\bibfnamefont {J.}~\bibnamefont {Kim}},\
  }\href {\doibase 10.1103/PhysRevB.92.054426} {\bibfield  {journal} {\bibinfo
  {journal} {Physical Review B}\ }\textbf {\bibinfo {volume} {92}},\ \bibinfo
  {pages} {054426} (\bibinfo {year} {2015})}\BibitemShut {NoStop}%
\bibitem [{\citenamefont {Zhou}\ \emph {et~al.}(2015)\citenamefont {Zhou},
  \citenamefont {Wang}, \citenamefont {Hou}, \citenamefont {Huang},
  \citenamefont {Lu},\ and\ \citenamefont {Wu}}]{Zhou2015}%
  \BibitemOpen
  \bibfield  {author} {\bibinfo {author} {\bibfnamefont {H.}~\bibnamefont
  {Zhou}}, \bibinfo {author} {\bibfnamefont {L.}~\bibnamefont {Wang}}, \bibinfo
  {author} {\bibfnamefont {Y.}~\bibnamefont {Hou}}, \bibinfo {author}
  {\bibfnamefont {Z.}~\bibnamefont {Huang}}, \bibinfo {author} {\bibfnamefont
  {Q.}~\bibnamefont {Lu}}, \ and\ \bibinfo {author} {\bibfnamefont
  {W.}~\bibnamefont {Wu}},\ }\href {\doibase 10.1038/ncomms9980} {\bibfield
  {journal} {\bibinfo  {journal} {Nature Communications}\ }\textbf {\bibinfo
  {volume} {6}},\ \bibinfo {pages} {1} (\bibinfo {year} {2015})}\BibitemShut
  {NoStop}%
\bibitem [{\citenamefont {Sheng}\ \emph {et~al.}(2018)\citenamefont {Sheng},
  \citenamefont {Feng}, \citenamefont {Zhou}, \citenamefont {Dong},
  \citenamefont {Xu}, \citenamefont {Cheng}, \citenamefont {Liu}, \citenamefont
  {Hou}, \citenamefont {Meng}, \citenamefont {Sun}, \citenamefont {Nakamura},
  \citenamefont {Tokura}, \citenamefont {Kawasaki},\ and\ \citenamefont
  {Lu}}]{Sheng2018}%
  \BibitemOpen
  \bibfield  {author} {\bibinfo {author} {\bibfnamefont {Z.}~\bibnamefont
  {Sheng}}, \bibinfo {author} {\bibfnamefont {Q.}~\bibnamefont {Feng}},
  \bibinfo {author} {\bibfnamefont {H.}~\bibnamefont {Zhou}}, \bibinfo {author}
  {\bibfnamefont {S.}~\bibnamefont {Dong}}, \bibinfo {author} {\bibfnamefont
  {X.}~\bibnamefont {Xu}}, \bibinfo {author} {\bibfnamefont {L.}~\bibnamefont
  {Cheng}}, \bibinfo {author} {\bibfnamefont {C.}~\bibnamefont {Liu}}, \bibinfo
  {author} {\bibfnamefont {Y.}~\bibnamefont {Hou}}, \bibinfo {author}
  {\bibfnamefont {W.}~\bibnamefont {Meng}}, \bibinfo {author} {\bibfnamefont
  {Y.}~\bibnamefont {Sun}}, \bibinfo {author} {\bibfnamefont {M.}~\bibnamefont
  {Nakamura}}, \bibinfo {author} {\bibfnamefont {Y.}~\bibnamefont {Tokura}},
  \bibinfo {author} {\bibfnamefont {M.}~\bibnamefont {Kawasaki}}, \ and\
  \bibinfo {author} {\bibfnamefont {Q.}~\bibnamefont {Lu}},\ }\href {\doibase
  10.1021/acsami.8b04057} {\bibfield  {journal} {\bibinfo  {journal} {ACS
  Applied Materials {\&} Interfaces}\ }\textbf {\bibinfo {volume} {10}},\
  \bibinfo {pages} {20136} (\bibinfo {year} {2018})}\BibitemShut {NoStop}%
\bibitem [{\citenamefont {Ueda}\ \emph {et~al.}(2005)\citenamefont {Ueda},
  \citenamefont {Katori}, \citenamefont {Mitamura}, \citenamefont {Goto},\ and\
  \citenamefont {Takagi}}]{Ueda2005}%
  \BibitemOpen
  \bibfield  {author} {\bibinfo {author} {\bibfnamefont {H.}~\bibnamefont
  {Ueda}}, \bibinfo {author} {\bibfnamefont {H.~A.}\ \bibnamefont {Katori}},
  \bibinfo {author} {\bibfnamefont {H.}~\bibnamefont {Mitamura}}, \bibinfo
  {author} {\bibfnamefont {T.}~\bibnamefont {Goto}}, \ and\ \bibinfo {author}
  {\bibfnamefont {H.}~\bibnamefont {Takagi}},\ }\href {\doibase
  10.1103/PhysRevLett.94.047202} {\bibfield  {journal} {\bibinfo  {journal}
  {Physical Review Letters}\ }\textbf {\bibinfo {volume} {94}},\ \bibinfo
  {pages} {2} (\bibinfo {year} {2005})}\BibitemShut {NoStop}%
\bibitem [{\citenamefont {Tokunaga}(2011)}]{Tokunaga2011}%
  \BibitemOpen
  \bibfield  {author} {\bibinfo {author} {\bibfnamefont {M.}~\bibnamefont
  {Tokunaga}},\ }\href {\doibase 10.1007/s11467-011-0203-2} {\bibfield
  {journal} {\bibinfo  {journal} {Frontiers of Physics}\ }\textbf {\bibinfo
  {volume} {7}},\ \bibinfo {pages} {386} (\bibinfo {year} {2011})}\BibitemShut
  {NoStop}%
\bibitem [{\citenamefont {Tao}\ \emph {et~al.}(2017)\citenamefont {Tao},
  \citenamefont {Singh}, \citenamefont {Rossi}, \citenamefont {Gerritsen},
  \citenamefont {Hendriksen}, \citenamefont {Khajetoorians}, \citenamefont
  {Christianen}, \citenamefont {Maan}, \citenamefont {Zeitler},\ and\
  \citenamefont {Bryant}}]{Tao2017}%
  \BibitemOpen
  \bibfield  {author} {\bibinfo {author} {\bibfnamefont {W.}~\bibnamefont
  {Tao}}, \bibinfo {author} {\bibfnamefont {S.}~\bibnamefont {Singh}}, \bibinfo
  {author} {\bibfnamefont {L.}~\bibnamefont {Rossi}}, \bibinfo {author}
  {\bibfnamefont {J.~W.}\ \bibnamefont {Gerritsen}}, \bibinfo {author}
  {\bibfnamefont {B.~L.~M.}\ \bibnamefont {Hendriksen}}, \bibinfo {author}
  {\bibfnamefont {A.~A.}\ \bibnamefont {Khajetoorians}}, \bibinfo {author}
  {\bibfnamefont {P.~C.~M.}\ \bibnamefont {Christianen}}, \bibinfo {author}
  {\bibfnamefont {J.~C.}\ \bibnamefont {Maan}}, \bibinfo {author}
  {\bibfnamefont {U.}~\bibnamefont {Zeitler}}, \ and\ \bibinfo {author}
  {\bibfnamefont {B.}~\bibnamefont {Bryant}},\ }\href {\doibase
  10.1063/1.4995372} {\bibfield  {journal} {\bibinfo  {journal} {Review of
  Scientific Instruments}\ }\textbf {\bibinfo {volume} {88}},\ \bibinfo {pages}
  {093706} (\bibinfo {year} {2017})}\BibitemShut {NoStop}%
\bibitem [{\citenamefont {Wiegers}\ \emph {et~al.}(2012)\citenamefont
  {Wiegers}, \citenamefont {Rook}, \citenamefont {den Ouden}, \citenamefont
  {Perenboom},\ and\ \citenamefont {Maan}}]{Wiegers2012}%
  \BibitemOpen
  \bibfield  {author} {\bibinfo {author} {\bibfnamefont {S.~A.~J.}\
  \bibnamefont {Wiegers}}, \bibinfo {author} {\bibfnamefont {J.}~\bibnamefont
  {Rook}}, \bibinfo {author} {\bibfnamefont {A.}~\bibnamefont {den Ouden}},
  \bibinfo {author} {\bibfnamefont {J.~A. A.~J.}\ \bibnamefont {Perenboom}}, \
  and\ \bibinfo {author} {\bibfnamefont {J.~C.}\ \bibnamefont {Maan}},\ }\href
  {\doibase 10.1109/TASC.2011.2177432} {\bibfield  {journal} {\bibinfo
  {journal} {IEEE Transactions on Applied Superconductivity}\ }\textbf
  {\bibinfo {volume} {22}},\ \bibinfo {pages} {4301504} (\bibinfo {year}
  {2012})}\BibitemShut {NoStop}%
\bibitem [{\citenamefont {Wiegers}\ \emph {et~al.}(2008)\citenamefont
  {Wiegers}, \citenamefont {Rook}, \citenamefont {Bird}, \citenamefont {Toth},
  \citenamefont {Bole}, \citenamefont {Perenboom},\ and\ \citenamefont
  {Maan}}]{Wiegers2008}%
  \BibitemOpen
  \bibfield  {author} {\bibinfo {author} {\bibfnamefont {S.~A.~J.}\
  \bibnamefont {Wiegers}}, \bibinfo {author} {\bibfnamefont {J.}~\bibnamefont
  {Rook}}, \bibinfo {author} {\bibfnamefont {M.~D.}\ \bibnamefont {Bird}},
  \bibinfo {author} {\bibfnamefont {J.}~\bibnamefont {Toth}}, \bibinfo {author}
  {\bibfnamefont {S.}~\bibnamefont {Bole}}, \bibinfo {author} {\bibfnamefont
  {J.~A. A.~J.}\ \bibnamefont {Perenboom}}, \ and\ \bibinfo {author}
  {\bibfnamefont {J.~C.}\ \bibnamefont {Maan}},\ }\href {\doibase
  10.1109/TASC.2008.920555} {\bibfield  {journal} {\bibinfo  {journal} {IEEE
  Transactions on Applied Superconductivity}\ }\textbf {\bibinfo {volume}
  {18}},\ \bibinfo {pages} {564} (\bibinfo {year} {2008})}\BibitemShut
  {NoStop}%
\bibitem [{\citenamefont {Giessibl}\ \emph {et~al.}(2011)\citenamefont
  {Giessibl}, \citenamefont {Pielmeier}, \citenamefont {Eguchi}, \citenamefont
  {An},\ and\ \citenamefont {Hasegawa}}]{Giessibl2011}%
  \BibitemOpen
  \bibfield  {author} {\bibinfo {author} {\bibfnamefont {F.~J.}\ \bibnamefont
  {Giessibl}}, \bibinfo {author} {\bibfnamefont {F.}~\bibnamefont {Pielmeier}},
  \bibinfo {author} {\bibfnamefont {T.}~\bibnamefont {Eguchi}}, \bibinfo
  {author} {\bibfnamefont {T.}~\bibnamefont {An}}, \ and\ \bibinfo {author}
  {\bibfnamefont {Y.}~\bibnamefont {Hasegawa}},\ }\href {\doibase
  10.1103/PhysRevB.84.125409} {\bibfield  {journal} {\bibinfo  {journal}
  {Physical Review B}\ }\textbf {\bibinfo {volume} {84}},\ \bibinfo {pages}
  {125409} (\bibinfo {year} {2011})}\BibitemShut {NoStop}%
\bibitem [{\citenamefont {Tortonese}\ \emph {et~al.}(1993)\citenamefont
  {Tortonese}, \citenamefont {Barrett},\ and\ \citenamefont
  {Quate}}]{tortonese1993}%
  \BibitemOpen
  \bibfield  {author} {\bibinfo {author} {\bibfnamefont {M.}~\bibnamefont
  {Tortonese}}, \bibinfo {author} {\bibfnamefont {R.~C.}\ \bibnamefont
  {Barrett}}, \ and\ \bibinfo {author} {\bibfnamefont {C.~F.}\ \bibnamefont
  {Quate}},\ }\href {\doibase 10.1063/1.108593} {\bibfield  {journal} {\bibinfo
   {journal} {Applied Physics Letters}\ }\textbf {\bibinfo {volume} {62}},\
  \bibinfo {pages} {834} (\bibinfo {year} {1993})}\BibitemShut {NoStop}%
\bibitem [{\citenamefont {Takahashi}\ \emph {et~al.}(2002)\citenamefont
  {Takahashi}, \citenamefont {Ando},\ and\ \citenamefont
  {Shirakawabe}}]{Takahashi2002}%
  \BibitemOpen
  \bibfield  {author} {\bibinfo {author} {\bibfnamefont {H.}~\bibnamefont
  {Takahashi}}, \bibinfo {author} {\bibfnamefont {K.}~\bibnamefont {Ando}}, \
  and\ \bibinfo {author} {\bibfnamefont {Y.}~\bibnamefont {Shirakawabe}},\
  }\href {\doibase 10.1016/S0304-3991(02)00083-9} {\bibfield  {journal}
  {\bibinfo  {journal} {Ultramicroscopy}\ }\textbf {\bibinfo {volume} {91}},\
  \bibinfo {pages} {63} (\bibinfo {year} {2002})}\BibitemShut {NoStop}%
\bibitem [{\citenamefont {Dukic}\ \emph {et~al.}(2015)\citenamefont {Dukic},
  \citenamefont {Adams},\ and\ \citenamefont {Fantner}}]{Dukic2015}%
  \BibitemOpen
  \bibfield  {author} {\bibinfo {author} {\bibfnamefont {M.}~\bibnamefont
  {Dukic}}, \bibinfo {author} {\bibfnamefont {J.~D.}\ \bibnamefont {Adams}}, \
  and\ \bibinfo {author} {\bibfnamefont {G.~E.}\ \bibnamefont {Fantner}},\
  }\href {\doibase 10.1038/srep16393} {\bibfield  {journal} {\bibinfo
  {journal} {Scientific Reports}\ }\textbf {\bibinfo {volume} {5}},\ \bibinfo
  {pages} {16393} (\bibinfo {year} {2015})}\BibitemShut {NoStop}%
\bibitem [{\citenamefont {Soghomonian}\ \emph {et~al.}(2000)\citenamefont
  {Soghomonian}, \citenamefont {Sabo}, \citenamefont {Powell}, \citenamefont
  {Murphy}, \citenamefont {Rosanske}, \citenamefont {Cross},\ and\
  \citenamefont {Schneider-Muntau}}]{Soghomonian2000}%
  \BibitemOpen
  \bibfield  {author} {\bibinfo {author} {\bibfnamefont {V.}~\bibnamefont
  {Soghomonian}}, \bibinfo {author} {\bibfnamefont {M.}~\bibnamefont {Sabo}},
  \bibinfo {author} {\bibfnamefont {A.}~\bibnamefont {Powell}}, \bibinfo
  {author} {\bibfnamefont {P.}~\bibnamefont {Murphy}}, \bibinfo {author}
  {\bibfnamefont {R.}~\bibnamefont {Rosanske}}, \bibinfo {author}
  {\bibfnamefont {T.~A.}\ \bibnamefont {Cross}}, \ and\ \bibinfo {author}
  {\bibfnamefont {H.~J.}\ \bibnamefont {Schneider-Muntau}},\ }\href {\doibase
  10.1063/1.1150707} {\bibfield  {journal} {\bibinfo  {journal} {Review of
  Scientific Instruments}\ }\textbf {\bibinfo {volume} {71}},\ \bibinfo {pages}
  {2882} (\bibinfo {year} {2000})}\BibitemShut {NoStop}%
\bibitem [{\citenamefont {Meyer}\ \emph {et~al.}(1989)\citenamefont {Meyer},
  \citenamefont {Silvera},\ and\ \citenamefont {Brandt}}]{Meyer1989}%
  \BibitemOpen
  \bibfield  {author} {\bibinfo {author} {\bibfnamefont {E.~S.}\ \bibnamefont
  {Meyer}}, \bibinfo {author} {\bibfnamefont {I.~F.}\ \bibnamefont {Silvera}},
  \ and\ \bibinfo {author} {\bibfnamefont {B.~L.}\ \bibnamefont {Brandt}},\
  }\href {\doibase 10.1063/1.1140636} {\bibfield  {journal} {\bibinfo
  {journal} {Review of Scientific Instruments}\ }\textbf {\bibinfo {volume}
  {60}},\ \bibinfo {pages} {2964} (\bibinfo {year} {1989})}\BibitemShut
  {NoStop}%
\bibitem [{\citenamefont {Inami}\ \emph {et~al.}(2006)\citenamefont {Inami},
  \citenamefont {Ohwada}, \citenamefont {Tsubota}, \citenamefont {Murata},
  \citenamefont {Matsuda}, \citenamefont {Nojiri}, \citenamefont {Ueda},\ and\
  \citenamefont {Murakami}}]{Inami2006}%
  \BibitemOpen
  \bibfield  {author} {\bibinfo {author} {\bibfnamefont {T.}~\bibnamefont
  {Inami}}, \bibinfo {author} {\bibfnamefont {K.}~\bibnamefont {Ohwada}},
  \bibinfo {author} {\bibfnamefont {M.}~\bibnamefont {Tsubota}}, \bibinfo
  {author} {\bibfnamefont {Y.}~\bibnamefont {Murata}}, \bibinfo {author}
  {\bibfnamefont {Y.~H.}\ \bibnamefont {Matsuda}}, \bibinfo {author}
  {\bibfnamefont {H.}~\bibnamefont {Nojiri}}, \bibinfo {author} {\bibfnamefont
  {H.}~\bibnamefont {Ueda}}, \ and\ \bibinfo {author} {\bibfnamefont
  {Y.}~\bibnamefont {Murakami}},\ }\href {\doibase 10.1088/1742-6596/51/1/114}
  {\bibfield  {journal} {\bibinfo  {journal} {Journal of Physics: Conference
  Series}\ }\textbf {\bibinfo {volume} {51}},\ \bibinfo {pages} {502} (\bibinfo
  {year} {2006})}\BibitemShut {NoStop}%
\bibitem [{\citenamefont {K{\"{u}}chler}\ \emph {et~al.}(2017)\citenamefont
  {K{\"{u}}chler}, \citenamefont {W{\"{o}}rl}, \citenamefont {Gegenwart},
  \citenamefont {Berben}, \citenamefont {Bryant},\ and\ \citenamefont
  {Wiedmann}}]{kuchler2017}%
  \BibitemOpen
  \bibfield  {author} {\bibinfo {author} {\bibfnamefont {R.}~\bibnamefont
  {K{\"{u}}chler}}, \bibinfo {author} {\bibfnamefont {A.}~\bibnamefont
  {W{\"{o}}rl}}, \bibinfo {author} {\bibfnamefont {P.}~\bibnamefont
  {Gegenwart}}, \bibinfo {author} {\bibfnamefont {M.}~\bibnamefont {Berben}},
  \bibinfo {author} {\bibfnamefont {B.}~\bibnamefont {Bryant}}, \ and\ \bibinfo
  {author} {\bibfnamefont {S.}~\bibnamefont {Wiedmann}},\ }\href {\doibase
  10.1063/1.4997073} {\bibfield  {journal} {\bibinfo  {journal} {Review of
  Scientific Instruments}\ }\textbf {\bibinfo {volume} {88}},\ \bibinfo {pages}
  {083903} (\bibinfo {year} {2017})}\BibitemShut {NoStop}%
\bibitem [{\citenamefont {Park}\ \emph {et~al.}(2009)\citenamefont {Park},
  \citenamefont {Graf}, \citenamefont {Murphy}, \citenamefont {Schmiedeshoff},\
  and\ \citenamefont {Tozer}}]{Park2009}%
  \BibitemOpen
  \bibfield  {author} {\bibinfo {author} {\bibfnamefont {J.-H.}\ \bibnamefont
  {Park}}, \bibinfo {author} {\bibfnamefont {D.}~\bibnamefont {Graf}}, \bibinfo
  {author} {\bibfnamefont {T.~P.}\ \bibnamefont {Murphy}}, \bibinfo {author}
  {\bibfnamefont {G.~M.}\ \bibnamefont {Schmiedeshoff}}, \ and\ \bibinfo
  {author} {\bibfnamefont {S.~W.}\ \bibnamefont {Tozer}},\ }\href {\doibase
  10.1063/1.3258143} {\bibfield  {journal} {\bibinfo  {journal} {Review of
  Scientific Instruments}\ }\textbf {\bibinfo {volume} {80}},\ \bibinfo {pages}
  {116101} (\bibinfo {year} {2009})}\BibitemShut {NoStop}%
\bibitem [{\citenamefont {Reid}\ \emph {et~al.}(2018)\citenamefont {Reid},
  \citenamefont {Shen}, \citenamefont {Maldonado}, \citenamefont {Chase},
  \citenamefont {Jal}, \citenamefont {Granitzka}, \citenamefont {Carva},
  \citenamefont {Li}, \citenamefont {Li}, \citenamefont {Wu}, \citenamefont
  {Vecchione}, \citenamefont {Liu}, \citenamefont {Chen}, \citenamefont
  {Higley}, \citenamefont {Hartmann}, \citenamefont {Coffee}, \citenamefont
  {Wu}, \citenamefont {Dakovski}, \citenamefont {Schlotter}, \citenamefont
  {Ohldag}, \citenamefont {Takahashi}, \citenamefont {Mehta}, \citenamefont
  {Hellwig}, \citenamefont {Fry}, \citenamefont {Zhu}, \citenamefont {Cao},
  \citenamefont {Fullerton}, \citenamefont {St{\"{o}}hr}, \citenamefont
  {Oppeneer}, \citenamefont {Wang},\ and\ \citenamefont
  {D{\"{u}}rr}}]{Reid2018}%
  \BibitemOpen
  \bibfield  {author} {\bibinfo {author} {\bibfnamefont {A.~H.}\ \bibnamefont
  {Reid}}, \bibinfo {author} {\bibfnamefont {X.}~\bibnamefont {Shen}}, \bibinfo
  {author} {\bibfnamefont {P.}~\bibnamefont {Maldonado}}, \bibinfo {author}
  {\bibfnamefont {T.}~\bibnamefont {Chase}}, \bibinfo {author} {\bibfnamefont
  {E.}~\bibnamefont {Jal}}, \bibinfo {author} {\bibfnamefont {P.~W.}\
  \bibnamefont {Granitzka}}, \bibinfo {author} {\bibfnamefont {K.}~\bibnamefont
  {Carva}}, \bibinfo {author} {\bibfnamefont {R.~K.}\ \bibnamefont {Li}},
  \bibinfo {author} {\bibfnamefont {J.}~\bibnamefont {Li}}, \bibinfo {author}
  {\bibfnamefont {L.}~\bibnamefont {Wu}}, \bibinfo {author} {\bibfnamefont
  {T.}~\bibnamefont {Vecchione}}, \bibinfo {author} {\bibfnamefont
  {T.}~\bibnamefont {Liu}}, \bibinfo {author} {\bibfnamefont {Z.}~\bibnamefont
  {Chen}}, \bibinfo {author} {\bibfnamefont {D.~J.}\ \bibnamefont {Higley}},
  \bibinfo {author} {\bibfnamefont {N.}~\bibnamefont {Hartmann}}, \bibinfo
  {author} {\bibfnamefont {R.}~\bibnamefont {Coffee}}, \bibinfo {author}
  {\bibfnamefont {J.}~\bibnamefont {Wu}}, \bibinfo {author} {\bibfnamefont
  {G.~L.}\ \bibnamefont {Dakovski}}, \bibinfo {author} {\bibfnamefont {W.~F.}\
  \bibnamefont {Schlotter}}, \bibinfo {author} {\bibfnamefont {H.}~\bibnamefont
  {Ohldag}}, \bibinfo {author} {\bibfnamefont {Y.~K.}\ \bibnamefont
  {Takahashi}}, \bibinfo {author} {\bibfnamefont {V.}~\bibnamefont {Mehta}},
  \bibinfo {author} {\bibfnamefont {O.}~\bibnamefont {Hellwig}}, \bibinfo
  {author} {\bibfnamefont {A.}~\bibnamefont {Fry}}, \bibinfo {author}
  {\bibfnamefont {Y.}~\bibnamefont {Zhu}}, \bibinfo {author} {\bibfnamefont
  {J.}~\bibnamefont {Cao}}, \bibinfo {author} {\bibfnamefont {E.~E.}\
  \bibnamefont {Fullerton}}, \bibinfo {author} {\bibfnamefont {J.}~\bibnamefont
  {St{\"{o}}hr}}, \bibinfo {author} {\bibfnamefont {P.~M.}\ \bibnamefont
  {Oppeneer}}, \bibinfo {author} {\bibfnamefont {X.~J.}\ \bibnamefont {Wang}},
  \ and\ \bibinfo {author} {\bibfnamefont {H.~A.}\ \bibnamefont {D{\"{u}}rr}},\
  }\href {\doibase 10.1038/s41467-017-02730-7} {\bibfield  {journal} {\bibinfo
  {journal} {Nature Communications}\ }\textbf {\bibinfo {volume} {9}},\
  \bibinfo {pages} {388} (\bibinfo {year} {2018})}\BibitemShut {NoStop}%
\bibitem [{\citenamefont {Poddar}\ \emph {et~al.}(2018)\citenamefont {Poddar},
  \citenamefont {de~Sa}, \citenamefont {Cai}, \citenamefont {Delannay},
  \citenamefont {Nysten}, \citenamefont {Piraux},\ and\ \citenamefont
  {Jonas}}]{Poddar2018}%
  \BibitemOpen
  \bibfield  {author} {\bibinfo {author} {\bibfnamefont {S.}~\bibnamefont
  {Poddar}}, \bibinfo {author} {\bibfnamefont {P.}~\bibnamefont {de~Sa}},
  \bibinfo {author} {\bibfnamefont {R.}~\bibnamefont {Cai}}, \bibinfo {author}
  {\bibfnamefont {L.}~\bibnamefont {Delannay}}, \bibinfo {author}
  {\bibfnamefont {B.}~\bibnamefont {Nysten}}, \bibinfo {author} {\bibfnamefont
  {L.}~\bibnamefont {Piraux}}, \ and\ \bibinfo {author} {\bibfnamefont {A.~M.}\
  \bibnamefont {Jonas}},\ }\href {\doibase 10.1021/acsnano.7b07389} {\bibfield
  {journal} {\bibinfo  {journal} {ACS Nano}\ }\textbf {\bibinfo {volume}
  {12}},\ \bibinfo {pages} {576} (\bibinfo {year} {2018})}\BibitemShut
  {NoStop}%
\bibitem [{\citenamefont {White}\ \emph {et~al.}(2011)\citenamefont {White},
  \citenamefont {Singh},\ and\ \citenamefont {Wahl}}]{White2011}%
  \BibitemOpen
  \bibfield  {author} {\bibinfo {author} {\bibfnamefont {S.~C.}\ \bibnamefont
  {White}}, \bibinfo {author} {\bibfnamefont {U.~R.}\ \bibnamefont {Singh}}, \
  and\ \bibinfo {author} {\bibfnamefont {P.}~\bibnamefont {Wahl}},\ }\href
  {\doibase 10.1063/1.3663611} {\bibfield  {journal} {\bibinfo  {journal}
  {Review of Scientific Instruments}\ }\textbf {\bibinfo {volume} {82}},\
  \bibinfo {pages} {1} (\bibinfo {year} {2011})}\BibitemShut {NoStop}%
\bibitem [{\citenamefont {Trainer}\ \emph {et~al.}(2017)\citenamefont
  {Trainer}, \citenamefont {Yim}, \citenamefont {McLaren},\ and\ \citenamefont
  {Wahl}}]{Trainer2017}%
  \BibitemOpen
  \bibfield  {author} {\bibinfo {author} {\bibfnamefont {C.}~\bibnamefont
  {Trainer}}, \bibinfo {author} {\bibfnamefont {C.~M.}\ \bibnamefont {Yim}},
  \bibinfo {author} {\bibfnamefont {M.}~\bibnamefont {McLaren}}, \ and\
  \bibinfo {author} {\bibfnamefont {P.}~\bibnamefont {Wahl}},\ }\href {\doibase
  10.1063/1.4995688} {\bibfield  {journal} {\bibinfo  {journal} {Review of
  Scientific Instruments}\ }\textbf {\bibinfo {volume} {88}},\ \bibinfo {pages}
  {093705} (\bibinfo {year} {2017})}\BibitemShut {NoStop}%
\bibitem [{\citenamefont {Tayler}\ and\ \citenamefont
  {Sakellariou}(2017)}]{Tayler2017}%
  \BibitemOpen
  \bibfield  {author} {\bibinfo {author} {\bibfnamefont {M.~C.}\ \bibnamefont
  {Tayler}}\ and\ \bibinfo {author} {\bibfnamefont {D.}~\bibnamefont
  {Sakellariou}},\ }\href {\doibase 10.1016/j.jmr.2017.03.001} {\bibfield
  {journal} {\bibinfo  {journal} {Journal of Magnetic Resonance}\ }\textbf
  {\bibinfo {volume} {277}},\ \bibinfo {pages} {143} (\bibinfo {year}
  {2017})}\BibitemShut {NoStop}%
\end{thebibliography}%
	
\end{document}